\documentclass[conference,a4paper]{IEEEtran}

\usepackage[utf8]{inputenc} 
\usepackage[T1]{fontenc}
\usepackage{url}              
\usepackage{cite}             

\usepackage[cmex10]{amsmath}  
\usepackage{mleftright}       
\mleftright                   

\usepackage{graphicx}         
\usepackage{booktabs}         

\usepackage{amsthm}
\usepackage{amssymb}
\usepackage{csquotes}
\usepackage{siunitx}
\usepackage{tikz}
\usepackage{pgfplots}
\usepackage{enumerate}
\usepackage{dsfont}
\usepackage{verbatim}
\usepackage[nolist]{acronym}

\usepackage{orcidlink} 

\newcommand{\given}{|}

\theoremstyle{definition}

\theoremstyle{remark}
\newtheorem*{prop}{Proposition}

\DeclareMathOperator*{\argmax}{arg\,max}

\definecolor{TUMpantone300}{RGB}{000,101,189}
\definecolor{TUMlightgreen}{RGB}{103,154,29}
\definecolor{TUMorangered}{RGB}{214,076,019}
\definecolor{TUMgray2}{RGB}{220,220,220}

\begin{document}

\begin{acronym}
    \acro{MI}{mutual information}
    \acro{GMI}{generalized mutual information}
    \acro{eBCH}{extended Bose-Chaudhuri-Hocquenghem}
    \acro{BDD}{bounded distance decoding}
    \acro{ML}{maximum likelihood}
    \acro{CN}{constraint node}
    \acro{APP}{a posteriori probability}
    \acro{LLR}{log-likelihood ratio}
    \acro{SPC}{single parity check}
    \acro{biAWGN}{binary-input additive white Gaussian noise}
    \acro{SNR}{signal-to-noise ratio}
    \acro{DE}{density evolution}
    \acro{BER}{bit error rate}
    \acro{LRB}{least reliable bit}
    \acro{GLDPC}{generalized low-density parity-check}
    \acro{LDPC}{low-density parity-check}
    \acro{KL}{Kullback-Leibler}
    \acro{BPSK}{binary phase-shift keying}
    \acro{ECPD-GMI}{extrinsic Chase-Pyndiah decoding with GMI-based soft-information post-processing}
    \acro{MCPD-GMI}{modified Chase-Pyndiah decoding with GMI-based soft-information post-processing}
    \acro{BP}{belief propagation}
    \acro{BICM}{bit-interleaved coded modulation}
\end{acronym}

\title{Soft-Information Post-Processing for \\Chase-Pyndiah Decoding Based on \\Generalized Mutual Information}

 \author{%
	   \IEEEauthorblockN{Andreas Straßhofer\IEEEauthorrefmark{1}\orcidlink{0009-0000-3572-0436},
		                     Diego Lentner\IEEEauthorrefmark{1}\orcidlink{0000-0001-6551-8925},
		                     Gianluigi Liva\IEEEauthorrefmark{2}\orcidlink{0000-0001-8657-2963},
		                     and Alexandre Graell~i~Amat\IEEEauthorrefmark{3}\orcidlink{0000-0002-5725-869X}}
	   \IEEEauthorblockA{\IEEEauthorrefmark{1}%
		                     Institute for Communications Engineering, 
		                     Technical University of Munich,
		                     Munich, Germany}
	   \IEEEauthorblockA{\IEEEauthorrefmark{2}%
	   						 Institute of Communications and Navigation
		                     German Aerospace Center (DLR),
		                     Wessling, Germany}
	   \IEEEauthorblockA{\IEEEauthorrefmark{3}%
		                     Chalmers University of Technology, 
		                     Gothenburg, Sweden}
	 }

\maketitle

\begin{abstract}
Chase-Pyndiah decoding is widely used for decoding product codes. However, this method is suboptimal and requires scaling the soft information exchanged during the iterative processing. In this paper, we propose a framework for obtaining the scaling coefficients based on maximizing the generalized mutual information. Our approach yields gains up to $0.11$~dB for product codes with two-error correcting extended BCH component codes over the binary-input additive white Gaussian noise channel compared to the original Chase-Pyndiah decoder with heuristically obtained coefficients. We also introduce an extrinsic version of the Chase-Pyndiah decoder and associate product codes with a turbo-like code ensemble to derive a Monte Carlo-based density evolution analysis. The resulting iterative decoding thresholds accurately predict the onset of the waterfall region.
\end{abstract}

\section{Introduction}
\label{sec:introduction}
Product codes \cite{Elias} when combined with Chase-Pyndiah decoding \cite{Pyndiah} feature an attractive performance-complexity tradeoff and are therefore widely used for applications with very high throughput requirements (e.g., \cite{Graell_i_Amat}). Chase-Pyndiah decoding is an efficient iterative soft-decision decoding algorithm that employs suboptimal Chase \cite{Chase} decoding of the component codes followed by an information combining step to generate the soft messages exchanged. To improve the finite-length performance, Pyndiah further proposed to scale the soft outputs by heuristically-determined parameters \cite{Pyndiah}. To the best of the authors' knowledge, no information-theoretically motivated method for optimizing these scaling coefficients has been reported to date. 

In this paper, we demonstrate that the \ac{GMI} is a suitable metric to determine the parameters to be used in the post-processing of the component code soft-output. The underlying idea is closely related to reconstructing soft information in min-sum decoding \cite{Lechner2} and coarsely quantized message passing \cite{Lechner}, \cite{Ben_Yacoub} for (generalized) \ac{LDPC} codes. The authors of \cite{Alvarado} derive a post-processing for a \ac{BICM} system using a cost function based on the \ac{GMI}.
Another related work \cite{Land} proposes to use the \ac{KL} divergence to assess the difference between \textit{accurate} and \textit{mismatched} reliability values. The use of the \ac{KL} divergence as optimization criterion showed promising results in the context of max-log \ac{APP} decoding of turbo codes.

The remainder of this paper is structured as follows. In Section~\ref{sec:preliminaries}, we review product codes and their decoding using the Chase-Pyndiah decoder. Section~\ref{sec:asymptotic-analysis} develops a method for analyzing a modified Chase-Pyndiah decoder in the asymptotic limit of large block lengths. In Section~\ref{sec:Chase-Pyndiah_with_GMI} we propose soft-information post-processing based on maximizing \ac{GMI}. Finally, we provide simulation results in Section~\ref{sec:numerical-results} and conclude the paper with Section~\ref{sec:conclusion}.

\section{Preliminaries} \label{sec:preliminaries}

\subsection{Channel Model}
We assume transmission over the \ac{biAWGN} channel $\mathsf{Y} = \mathsf{X} + \mathsf{Z}$, where $\mathsf{X} \in \{ +1,-1 \}$ and $\mathsf{Z} \sim \mathcal{N} ( 0, \sigma^2 )$. The channel quality is given as the \ac{SNR} $E_\text{b}/N_0 = \left( 2 R \sigma^2 \right)^{-1}$, where  $R$ is the  code rate. Given a channel output $y$, we compute the channel \ac{LLR} as
\begin{equation} \label{eq:channel_LLR}
    l^\text{ch} = \ln \left( \frac{p_{\mathsf{Y} \given \mathsf{X}} (y \given +1)}{p_{\mathsf{Y} \given \mathsf{X}} (y \given -1)} \right) = \frac{2}{\sigma^2} y\,.
\end{equation}

\subsection{Product Codes}

Product codes are serially concatenated codes, whose encoding is given by the following procedure. First,  the message bits are arranged into a $k \times k$ matrix. Then, each row is encoded using a row component code. Finally, each column of the resulting matrix is encoded by a column component code. In this paper, we focus on the practical case of identical row and column component codes.
In particular, we consider component codes with parameters $(n,k,d)$, where $n$ is the block length, $k$ the dimension, and $d$ the minimum Hamming distance of the code. Then, the parameters of the product code are $(n^2,k^2,d^2)$ and its rate $k^2/n^2$.

\subsection{Chase-Pyndiah Decoding}
\label{subsec:Chase-Pyndiah}

In this subsection we review Chase-Pyndiah decoding as in \cite{Pyndiah}. The decoder iterates between decoding the row and column component codes, where rows first or columns first can be chosen arbitrarily. In each half iteration $\ell = 1,\dots , \ell_\text{max}$, i.e., either row or column decoding, each component decoder processes an incoming message vector $\boldsymbol{l}^\text{in}$ of length $n$ as follows.

First, apply Chase decoding to $\boldsymbol{l}^\text{in}$. Herefore, obtain the hard-decision vector $\boldsymbol{r}=(r_1,\ldots,r_n)$, where
\begin{equation}
	r_i = 
	\begin{cases}
		0 & \text{if } l_i^\text{in} \geq 0\\
		1 & \text{otherwise.}
	\end{cases}
\end{equation}
From $\boldsymbol{l}^\text{in}$, identify the $p$ positions of lowest reliability $\given l_i^\text{in} \given$. From $\boldsymbol{r}$ generate a list of test words by the following bit flip procedure. Flip any single bit among the $p$ \acp{LRB}, giving $p$ test words. Then, flip any two bits to get $\binom{p}{2}$ more test words. Continue until the last test word, which is found by flipping all $p$ \acp{LRB}. Next,  decode the $2^p$ test words (including $\boldsymbol{r}$ itself) using \ac{BDD} to form the candidate list $\mathcal{L}$ consisting of all unique codewords resulting from successful \ac{BDD} attempts.

Second, for $i =1,\dots ,n$, search for the following two codewords in $\mathcal{L}$: One with bit value zero at position $i$ and highest likelihood given $\boldsymbol{l}^\text{in}$, and another one with bit value one at position $i$ and highest likelihood given $\boldsymbol{l}^\text{in}$. Clearly, one of them is the \ac{ML} decision among the codewords of the candidate list. The other codeword is referred to as the \textit{alternative codeword}. Generate the soft output as
\begin{equation} \label{eq:output}
	w_i = 
	\begin{cases}
		\frac{1}{2} d_i \sum\limits_{k \neq i} ( d_k - \bar{x}_k^{(i)} ) l_k^\text{in} & \text{if } \bar{\boldsymbol{x}}^{(i)} \text{ exists}\\
		d_i & \text{otherwise}
	\end{cases}
\end{equation}
where $\boldsymbol{d}$ and $\bar{\boldsymbol{x}}^{(i)}$ are the modulated \ac{ML} codeword and the modulated alternative codeword, respectively. In this work, we only consider values for $p$ that allow omitting the case of an empty candidate list.

Third, collect the soft outputs of all component codes in the matrix $\boldsymbol{W}$. Let $\mathcal{I}$ be a set of index pairs $(i,j)$ for which a component decoder found an alternative codeword. Then, Chase-Pyndiah decoding post-processes the soft output as
\begin{equation} \label{eq:w_to_m}
	v_{i,j} =
	\begin{cases}
		\alpha \cdot \left( \text{avg} \left( \given \boldsymbol{W}_{\mathcal{I}} \given \right) \right)^{-1} \cdot w_{i,j} & \text{if } \bar{\boldsymbol{x}}^{(i,j)} \text{ exists}\\
		\alpha \cdot \left( \text{avg} \left( \given \boldsymbol{W}_{\mathcal{I}} \given \right) \right)^{-1} \cdot \beta w_{i,j} & \text{otherwise}
	\end{cases}
\end{equation}
where $|\cdot|$ denotes the element-wise absolute value and $\text{avg} \left( \cdot \right)$ is the arithmetic average over all entries of a matrix. $\alpha$ and $\beta$ are the half iteration-dependent, heuristic scaling factors provided in \cite{Pyndiah}. The values of $\alpha$ and $\beta$ in the first eight half iterations are $0.1$, $0.3$, $0.5$, $0.7$, $0.9$, $1$, $1$, $1$ and $0.2$, $0.4$, $0.6$, $0.8$, $1$, $1$, $1$, $1$, respectively. In the subsequent half iterations, both $\alpha$ and $\beta$ are typically set to one as well.

Finally, let $\boldsymbol{L}^\text{ch}$ denote the channel \ac{LLR} matrix whose entries are computed according to \eqref{eq:channel_LLR}. The matrix of outgoing messages is then computed as (see \cite{Pyndiah})
\begin{equation} \label{eq:combining}
	\boldsymbol{L}^\text{out} = \left( \text{avg} \left( \given \boldsymbol{L}^\text{ch} \given \right) \right)^{-1} \cdot \boldsymbol{L}^\text{ch} + \boldsymbol{V}
\end{equation}
where the entries of the matrix $\boldsymbol{V}$ are the post-processed outputs obtained by \eqref{eq:w_to_m}. The output $\boldsymbol{L}^\text{out}$ of half iteration $\ell$ forms then the input $\boldsymbol{L}^\text{in}$ for the following half iteration $\ell+1$. 
Note that in the first half iteration $\boldsymbol{L}^\text{in}$ is initialized with the first term in \eqref{eq:combining}, i.e., the normalized channel \ac{LLR} matrix. 

Decoding terminates after $\ell_\text{max}$ half iterations. Then, use the modulated \ac{ML} codeword $\boldsymbol{d}$ of each component decoder obtained in the last half iteration to estimate the transmitted code matrix.

\subsection{Generalized Mutual Information}
\label{subsec:GMI_def}

Let $p_{\mathsf{Y} \given \mathsf{X}}$ be the transition probability function of a channel from $\mathsf{X}$ to $\mathsf{Y}$. Sometimes, e.g., due to complexity limitations, the receiver relies on a mismatched channel model $q_{\mathsf{Y} \given \mathsf{X}}$ and outputs the codeword that maximizes $q_{\mathsf{Y} \given \mathsf{X}}(x\given y)$ for a given channel output $y$. Using this decoder, an achievable rate is the $s$-\ac{GMI} \cite{Kaplan}
\begin{equation}
	I_s \left( \mathsf{X};\mathsf{Y} \right) = \mathds{E}_{p_\mathsf{XY}} \left\{ \log_2 \left( \frac{q_{\mathsf{Y} \given \mathsf{X}} (\mathsf{Y} \given \mathsf{X} )^s}{\sum\limits_{x' \in \mathcal{X}} p_\mathsf{X}(x') q_{\mathsf{Y} \given \mathsf{X}} (\mathsf{Y} \given x')^s} \right) \right\}
\end{equation}
with $s \geq 0$. We may simplify our considerations by setting $s=1$ since the $s$-\ac{GMI} is an achievable rate for any parameter $s$. An important upper bound to $I_1(\mathsf{X};\mathsf{Y})$ is the \ac{MI} $I(\mathsf{X};\mathsf{Y})$, which is achieved if $p_{\mathsf{Y} \given \mathsf{X}}=q_{\mathsf{Y} \given \mathsf{X}}$. In the following, we refer to the $1$-\ac{GMI} simply as \ac{GMI}.

\section{Asymptotic Analysis} \label{sec:asymptotic-analysis}

In this section, we provide a Monte-Carlo \ac{DE} of product codes under Chase-Pyndiah decoding. In the following, we consider a product code as an instance of a turbo-like code ensemble \cite{Moloudi} and apply changes to the original Chase-Pyndiah decoder as in Section~\ref{subsec:Chase-Pyndiah} such that the asymptotic analysis becomes applicable.

\subsection{Turbo-like Code Ensemble}
\label{subsec:tlc-ensemble}
\begin{figure}
	\centering
	\includegraphics[scale=0.8]{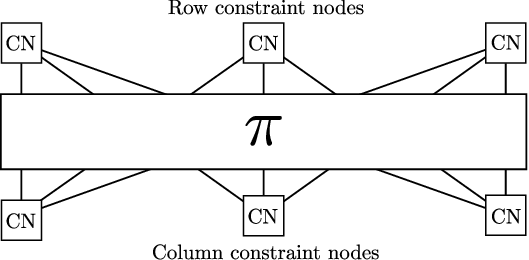}
	\caption{Turbo-like code ensemble for SPC component codes of length $n=3$.}
	\label{fig:simplified_Tanner_graph}
\end{figure}

A product code with $(n,k,d)$ component codes can be represented by a bipartite graph. Each of the $2n$ \acp{CN} corresponds to the code constraints imposed by one component code. The edges are associated with the $n^2$ code bits. Two \acp{CN} are connected by an edge if the associated code bit is part of the code constraints of the corresponding component codes. Finally, partitioning \acp{CN} into row \acp{CN} and column \acp{CN} leads to a bipartite graph. The deterministic product code structure defines a specific connection of row and column \acp{CN}. The set of codes defined by all possibilities to connect the row and column \acp{CN} gives the turbo-like code ensemble of length $n^2$. In Fig.~\ref{fig:simplified_Tanner_graph}, we use the edge interleaver $\pi$ to visualize the turbo-like code ensemble that contains the $(9,4,4)$ product code built upon $(3,2,2)$ \ac{SPC} component codes as an instance.
The design rate of the code ensemble is given by
\begin{equation}
	R_\text{d} = \frac{n^2 - 2 n \left( n - k \right)}{n^2} = 2 \sqrt{R} - 1
\end{equation}
where $R$ is the rate of the product code. We can now generate a sequence of code ensembles with increasing length by adding further \acp{CN} and edges. We remark that Chase-Pyndiah decoding can be seen as a message passing over the edges of the bipartite graph of the code. Then, each \ac{CN} processes incoming messages $\boldsymbol{l}^\text{in}$ into outgoing messages $\boldsymbol{l}^\text{out}$.

\subsection{Extrinsic Chase-Pyndiah Decoding} \label{subsec:Extrinsic_Chase_Pyndiah_Decoding}
\Ac{DE} analysis requires \textit{extrinsic message passing} over the edges of the bipartite graph \cite{Jian}, which is not the case for Chase-Pyndiah decoding. We therefore apply the following changes to the decoding process as described in Section~\ref{subsec:Chase-Pyndiah} to make it extrinsic \cite{Sheikh1}.
\begin{enumerate}
	\item Every component decoder computes each $w_i$, $i=1,\dots,n$, by replacing the $i$-th input message with the channel \ac{LLR}, i.e., we set $l_i^\text{in} = l_i^\text{ch}$. Note that we may obtain a different candidate list for each $i$.
	\item We set $v_i = w_i$, i.e., we omit the post-processing in \eqref{eq:w_to_m}.
	\item In \eqref{eq:combining}, the matrix $\boldsymbol{L}^\text{ch}$ is not normalized.
    \item We compute the a posteriori \ac{LLR} as $l_i^\text{APP} = v_i + l_i^\text{ch} + l_i^\text{in}$ and obtain the final decision as
        \begin{equation}
            \hat{c}_i =
            \begin{cases}
                0 & \text{if } l_i^\text{APP} \geq 0 \\
                1 & \text{otherwise.}
            \end{cases}
        \end{equation}
\end{enumerate}
We refer to this modified algorithm as \emph{extrinsic Chase-Pyndiah decoding}. Note that due to the modification in Step 1, computing the soft output $\boldsymbol{w}$ is $n$ times more complex than for original Chase-Pyndiah decoding. However, the extrinsic Chase-Pyndiah decoder is only used as a proxy to enable the \ac{DE} analysis of the turbo-like code ensemble defined in Section~\ref{subsec:tlc-ensemble} under Chase-Pyndiah decoding.

\subsection{Monte-Carlo Density Evolution}

Assume the all-zero codeword is transmitted over the \ac{biAWGN} channel with \ac{SNR} $E_\text{b}/N_0$. Then, the channel \acp{LLR} are Gaussian distributed with mean $\mu_\text{ch} = 4 R_\text{d} \left( E_\text{b}/N_0 \right)$ and variance $\sigma_\text{ch}^2 = 2 \mu_\text{ch}$. We want to track the distribution of the post-processed output $p_\mathsf{V}^{(\ell)}$ and the distribution of the incoming messages $p_{\mathsf{L}^\text{in}}^{(\ell)}$ throughout the half iterations $\ell = 1, 2, \dots $. We do so using a Monte-Carlo simulation. The ensemble iterative decoding threshold $\left( E_\text{b}/N_0 \right)^\star$ is defined as the lowest \ac{SNR} for which the probability $\text{Pr} \left[ \mathsf{V}^{(\ell)} < 0 \right]$ tends to zero as the number of half iterations grows large. Note that a Monte-Carlo simulation can only approximate the iterative decoding threshold by processing a finite number of samples with a finite number of half iterations $\ell_\text{max}$. We permute the incoming messages before each half iteration to mitigate any dependencies introduced in previous half iterations \cite{MacKay}.

\section{Extrinsic Chase-Pyndiah Decoding with GMI-based Soft-Information Post-Processing} \label{sec:Chase-Pyndiah_with_GMI}

\begin{figure}
	\centering
	\includegraphics[scale=0.9]{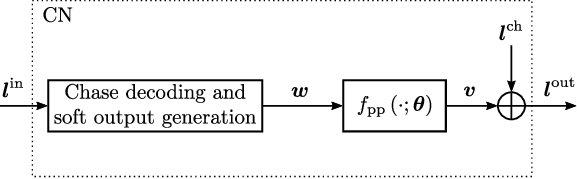}
	\caption{Chase decoding and soft output generation followed by post-processing and addition of the channel \ac{LLR}.}
	\label{fig:model}
\end{figure}

\Ac{BP} decoding of product codes employs optimal \ac{APP} decoding of the component codes at the \acp{CN} and  iteratively exchanges extrinsic \acp{LLR} between row and column decoders. The outputs $w_i$ of \eqref{eq:output}, however, can only be considered as \emph{approximate} \ac{LLR} values since they violate the \emph{consistency condition} \cite{Richardson}.
Pyndiah therefore proposed to post-process $\boldsymbol{w}$ as in \eqref{eq:w_to_m} and \eqref{eq:combining} to improve the performance. The post-processed output $v_i$ is then in general considered to be \emph{closer} to the true \ac{LLR} values. Clearly, the same argument applies also to extrinsic Chase-Pyndiah decoding as introduced in Section~\ref{subsec:Extrinsic_Chase_Pyndiah_Decoding}.

We propose to substitute this heuristic post-processing by a post-processing function $v_i = f_\text{pp} \left( w_i; \boldsymbol{\theta} \right)$ based on the GMI and parametrized by 
$\boldsymbol{\theta}= \left( \gamma , \delta \right)$. Fig.~\ref{fig:model} depicts the model under investigation, consisting of Chase decoding and soft output generation followed by the post-processing function and addition of the channel \ac{LLR}.

For simplicity, we focus on functions of the form
\begin{equation} \label{eq:postprocessing_function}
	f_\text{pp} \left( w_i; \boldsymbol{\theta} \right) = 
	\begin{cases}
		\gamma w_i & \text{if } \bar{\boldsymbol{x}}^{(i)} \text{ exists} \\
		\delta w_i & \text{otherwise.}
	\end{cases}
\end{equation}
Similar to $\alpha$ and $\beta$ in \eqref{eq:w_to_m}, $\gamma$ and $\delta$ are $\ell$-dependent parameters that can be precomputed offline.

Under a mismatched-decoding perspective, we can  rephrase the problem as follows. Consider an arbitrary outgoing message $l^\text{out}$ after $\ell$ half iterations. The decoder in the $\ell+1$-th half iteration wrongly \emph{interprets} this continuous value as an actual \ac{LLR}, i.e., it imposes a mismatched model $q_{\mathsf{L}^\text{out}\given\mathsf{X}}$ that satisfies 
\begin{equation} \label{eq:L-value_auxiliary}
	l^\text{out} = \ln \left( \frac{q_{\mathsf{L}^\text{out} \given \mathsf{X}} \left( l^\text{out} \given +1 \right)}{q_{\mathsf{L}^\text{out} \given \mathsf{X}} \left( l^\text{out} \given -1 \right)} \right)
\end{equation}
for any realization $l^\text{out}$.
An achievable rate for this mismatched setup is then the \ac{GMI} (see Section~\ref{subsec:GMI_def})
\begin{align}
    I_1 \left( \mathsf{X}; \mathsf{L}^\text{out} \right) &= \mathds{E}_{p_{\mathsf{L}^\text{out} \mathsf{X}}} \left\{ \log_2 \left( \frac{q_{\mathsf{L}^\text{out} \given \mathsf{X}} \left( \mathsf{L}^\text{out} \given \mathsf{X} \right)}{\sum\limits_{x' \in \mathcal{X}} p_\mathsf{X} (x') q_{\mathsf{L}^\text{out} \given \mathsf{X}} \left( \mathsf{L}^\text{out} \given x' \right)} \right) \right\} \, .
\end{align}
The following proposition shows that $I_1 \left( \mathsf{X}; \mathsf{L}^\text{out} \right)$ remarkably does not depend on the actual choice of $q_{\mathsf{L}^\text{out}\given \mathsf{X}}$ as long as it satisfies \eqref{eq:L-value_auxiliary}.

\begin{figure}
	\centering
	\begin{tikzpicture} 
		\begin{semilogyaxis}[label style={font=\footnotesize}, ticklabel style = {font=\footnotesize}, xlabel={$E_\text{b}/N_0$ [dB]}, ylabel=BER, xmin=1.5, xmax=4, ymin=3e-6, ymax=7e-2, grid=both, grid style={TUMgray2}, width=\columnwidth, height=6.5cm]
			
			\addplot[line width=1, TUMorangered, solid, mark=triangle*, mark options={fill=white}] coordinates {(1.7,0.06314) (1.8,0.0337) (1.9,0.007608) (2,0.001666) (2.1,0.00039) (2.175,0.0001441) (2.2, 9.995e-05) (2.25, 2.727e-05)};
			\label{line:64};
			
			\addplot[line width=1, TUMlightgreen, solid, mark=square*, mark options={fill=white}] coordinates {(2.6,0.03794) (2.7,0.02884) (2.75,0.01393) (2.8,0.002858) (2.85,0.0003924) (2.9,8.5e-05) (2.95,1.104e-5)};
			\label{line:128};
			
			\addplot[line width=1, TUMpantone300, mark=*, mark options={fill=white}] coordinates {(3.5, 0.01997) (3.6, 0.012665) (3.65, 0.00289187) (3.675, 5.326e-4) (3.7, 6.297e-5) (3.7125, 1.513e-5) (3.725, 4.669e-6)};
			\label{line:256};
			
			\addplot[line width=1, TUMorangered, mark=triangle*, mark options={solid,fill=white}, dashed] coordinates {(1.7,0.0604) (1.8,0.03865) (1.9,0.01925) (2,0.007718) (2.1,0.002393) (2.2,0.000451) (2.25, 0.0001764) (2.3, 5.907e-05) (2.35, 1.766e-5)};
			\label{line:original_64};
			
			\addplot[line width=1, TUMlightgreen, mark=square*, mark options={solid,fill=white}, dashed] coordinates {(2.6,0.0529) (2.7,0.03452) (2.8,0.01419) (2.85,0.006415) (2.9,0.002517) (2.95,0.0006221) (3,0.0001149) (3.05,1.653e-5)};
			\label{line:original_128};
			
			\addplot[line width=1, TUMpantone300, mark=*, mark options={solid,fill=white}, dashed] coordinates {(3.5, 0.03227) (3.6, 0.02178) (3.65, 0.01454) (3.7, 0.006971) (3.75,0.00127) (3.8, 0.0001188) (3.825, 1.006e-05)};
			\label{line:original_256};
			
		\end{semilogyaxis}
	\end{tikzpicture}
	\caption{BER performance of the $(64^2,51^2,6^2)$ (\ref{line:64}), $(128^2,113^2,6^2)$ (\ref{line:128}) and $(256^2, 239^2, 6^2)$ (\ref{line:256}) product codes under MCPD-GMI with $p=5$ after $20$ half iterations. Results for original Chase-Pyndiah decoding (\ref{line:original_64}, \ref{line:original_128}, \ref{line:original_256}) are provided for comparison.}
	\label{fig:num_res_1}
\end{figure}
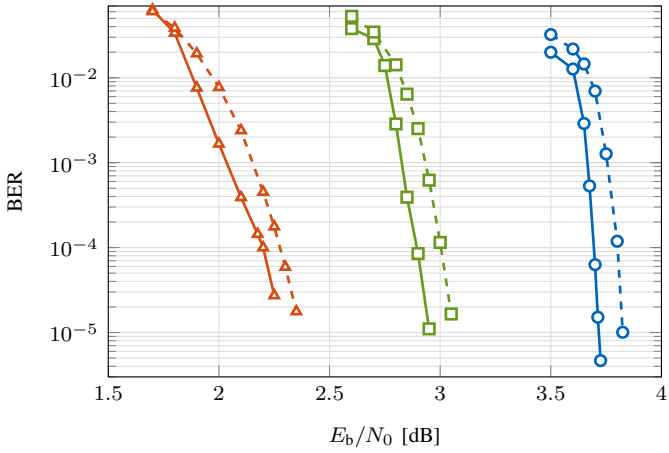

\begin{prop}
The GMI $I_1 \left( \mathsf{X}; \mathsf{L}^\text{out} \right)$ is given by
		\begin{equation}
		I_1 \left( \mathsf{X}; \mathsf{L}^\text{out} \right) = 1 - \mathds{E} \left\{ \log_2 \left[ 1 + \exp \left( - f_\text{pp} \left( \mathsf{W}; \boldsymbol{\theta} \right) - \mathsf{L}^\text{ch} \right) \right] \right\}\,,
		\label{eq:GMI_to_maximize}
	\end{equation}
	where the expectation is with respect to $p_{\mathsf{W} \mathsf{L}^\text{ch} \given \mathsf{X} = +1}$.
	
	\begin{proof}
        We have
		\begin{align}
				&I_1 \left( \mathsf{X}; \mathsf{L}^\text{out} \right) 
                \overset{(a)}{=} 1 + \mathds{E}_{p_{\mathsf{L}^\text{out} \mathsf{X}}} \left\{ \log_2 \left( \frac{q_{\mathsf{L}^\text{out} \given \mathsf{X}} \left( \mathsf{L}^\text{out} \given \mathsf{X} \right)}{\sum\limits_{x' \in \mathcal{X}} q_{\mathsf{L}^\text{out} \given \mathsf{X}} \left( \mathsf{L}^\text{out} \given x' \right)} \right) \right\} \nonumber\\
                &\qquad\overset{(b)}{=} 1 - \mathds{E}_{p_{\mathsf{L}^\text{out} \mathsf{X}}} \left\{ \log_2 \left( 1 + \left( \frac{q_{\mathsf{L}^\text{out} \given \mathsf{X}} \left( \mathsf{L}^\text{out} \given +1 \right)}{q_{\mathsf{L}^\text{out} \given \mathsf{X}} \left( \mathsf{L}^\text{out} \given -1 \right)} \right)^{-\mathsf{X}} \right)  \right\} \nonumber\\
                &\qquad\overset{(c)}{=} 1 - \mathds{E}_{p_{\mathsf{L}^\text{out} \mathsf{X}}} \left\{ \log_2 \left( 1 + \exp \left( - \mathsf{L}^\text{out} \mathsf{X} \right) \right)  \right\} \nonumber\\
				&\qquad\overset{(d)}{=} 1 - \mathds{E}_{p_{\mathsf{L}^\text{out} \given \mathsf{X} = +1}} \left\{ \log_2 \left( 1 + \exp \left( - \mathsf{L}^\text{out} \right) \right) \right\}
		\end{align}
        where $(a)$ and $(b)$ follow from $\mathsf{X} \in \left\{ -1, +1 \right\}$ being uniformly distributed; $(c)$ exploits that $q_{\mathsf{L}^\text{out}\given \mathsf{X}}$ must satisfy \eqref{eq:L-value_auxiliary} and $(d)$ uses the symmetry condition $p_{\mathsf{L}^\text{out} \given \mathsf{X}} \left( l^\text{out} \given -x \right) = p_{\mathsf{L}^\text{out} \given \mathsf{X}} \left( -l^\text{out} \given x \right)$.
		Resolving the output \ac{LLR} as given by our model as $\mathsf{L}^\text{out} = f_\text{pp} \left( \mathsf{W}; \boldsymbol{\theta} \right) + \mathsf{L}^\text{ch}$ completes the proof.
	\end{proof}
 
\end{prop}

Our key idea is to optimize the parameters $\boldsymbol{\theta}^\star = \left( \gamma^\star , \delta^\star \right)$ of the post-processing function \eqref{eq:postprocessing_function} so that  the \ac{GMI} \eqref{eq:GMI_to_maximize} is maximized, i.e.,
\begin{equation} \label{eq:optimal_parameters}
	\boldsymbol{\theta}^\star = \argmax_{\boldsymbol{\theta}} I_1 \left( \mathsf{X} ; \mathsf{L}^\text{out} \right)\,.
\end{equation}

We refer to the extrinsic Chase-Pyndiah decoding where the output $w$ is post-processed by a function with parameters defined by \eqref{eq:optimal_parameters} as \ac{ECPD-GMI}.

\section{Numerical Results} \label{sec:numerical-results}

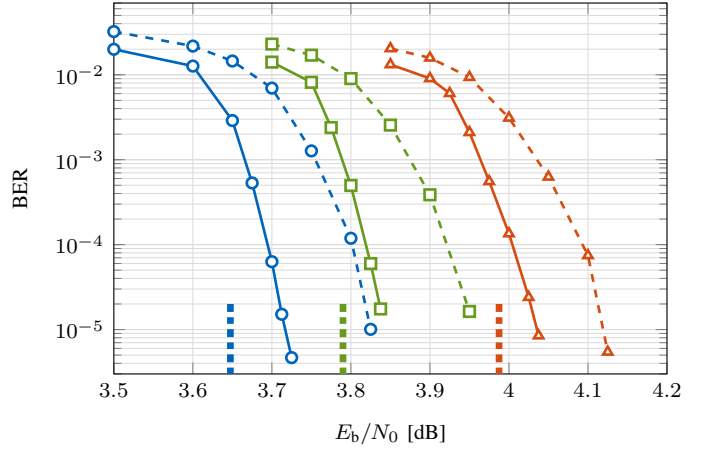
\begin{figure}
	\centering
	\begin{tikzpicture}
		\begin{semilogyaxis}[label style={font=\footnotesize}, ticklabel style = {font=\footnotesize}, xlabel={$E_\text{b}/N_0$ [dB]}, ylabel=BER, xmin=3.5, xmax=4.2, ymin=3e-6, ymax=7e-2, grid=both, grid style={TUMgray2}, width=\columnwidth, height=6.5cm]
			
			\addplot[line width=1, TUMorangered, mark=triangle*, mark options={fill=white}] coordinates {(3.85, 0.01321) (3.9, 0.009062) (3.925, 0.006048) (3.95, 0.0021) (3.975, 0.0005558) (4, 0.0001348) (4.025, 2.402e-05) (4.0375, 8.472e-06)};
			\label{line:p3};
			
			\addplot[line width=1, TUMlightgreen, mark=square*, mark options={fill=white}] coordinates { (3.7,0.0141) (3.75,0.008157) (3.775,0.002393) (3.8,0.0004967) (3.825,5.984e-05) (3.8375,1.749e-05)};
			\label{line:p4};
			
			\addplot[line width=1, TUMpantone300, mark=*, mark options={fill=white}] coordinates {(3.5, 0.01997) (3.6, 0.012665) (3.65, 0.00289187) (3.675, 5.326e-4) (3.7, 6.297e-5) (3.7125, 1.513e-5) (3.725, 4.669e-6)};
			\label{line:p5}
			
			\addplot[line width=1, TUMorangered, mark=triangle*, mark options={solid,fill=white}, dashed] coordinates {(3.85,0.0204) (3.9,0.01583) (3.95,0.009368) (4,0.003108) (4.05,0.0006274) (4.1,7.452e-05) (4.125,5.45e-06)};
			\label{line:original_p3};
			
			\addplot[line width=1, TUMlightgreen, mark=square*, mark options={solid,fill=white}, dashed] coordinates {(3.7,0.02298) (3.75,0.01704) (3.8,0.009023) (3.85,0.002555) (3.9,0.0003864) (3.95,1.632e-05) (4,0)};
			\label{line:original_p4};
			
			\addplot[line width=1, TUMpantone300, mark=*, mark options={solid,fill=white}, dashed] coordinates {(3.5, 0.03227) (3.6, 0.02178) (3.65, 0.01454) (3.7, 0.006971) (3.75,0.00127) (3.8, 0.0001188) (3.825, 1.006e-05)};
			\label{line:original_p5};
			
			\addplot[line width=2.5, TUMpantone300, no marks, dashdotted] coordinates {(3.6475, 2e-5) (3.6475, 1e-8)};
			\addplot[line width=2.5, TUMlightgreen, no marks, dashdotted] coordinates {(3.79, 2e-5) (3.79, 1e-8)};
			\addplot[line width=2.5, TUMorangered, no marks, dashdotted] coordinates {(3.9875, 2e-5) (3.9875, 1e-8)};
			
		\end{semilogyaxis}
	\end{tikzpicture}
	\caption{BER performance of the $(256^2, 239^2, 6^2)$ product code under MCPD-GMI with $p=5$ (\ref{line:p5}), $p=4$ (\ref{line:p4}) and $p=3$ (\ref{line:p3}). Results for original Chase-Pyndiah decoding (\ref{line:original_p5}, \ref{line:original_p4}, \ref{line:original_p3}) are provided for comparison. Vertical lines indicate iterative decoding thresholds computed using \ac{ECPD-GMI}.}
	\label{fig:num_res_2}
\end{figure}

\begin{table*}
	\centering
	\caption{Optimal parameters $\left( \gamma^\star, \delta^\star \right)$ for the $(256^2,239^2,6^2)$ product code decoded with $p=5$ at an SNR of $3.7$ dB.}
	\begin{tabular}{|b{0.3cm}|b{0.4cm}|b{0.4cm}|b{0.4cm}|b{0.4cm}|b{0.4cm}|b{0.4cm}|b{0.4cm}|b{0.4cm}|b{0.4cm}|b{0.4cm}|b{0.4cm}|b{0.4cm}|b{0.4cm}|b{0.4cm}|b{0.4cm}|b{0.4cm}|b{0.4cm}|b{0.4cm}|b{0.4cm}|b{0.4cm}|}
		\hline
		$\ell$ & $1$ & $2$ & $3$ & $4$ & $5$ & $6$ & $7$ & $8$ & $9$ & $10$ & $11$ & $12$ & $13$ & $14$ & $15$ & $16$ & $17$ & $18$ & $19$ & $20$\\
		\hline
		$\gamma^\star$ & $0.16$ & $0.20$ & $0.21$ & $0.22$ & $0.24$ & $0.25$ & $0.27$ & $0.28$ & $0.30$ & $0.33$ & $0.35$ & $0.37$ & $0.39$ & $0.42$ & $0.43$ & $0.43$ & $0.43$ & $0.42$ & $0.41$ & $0.39$\\
		\hline
		$\delta^\star$ & $0.15$ & $0.10$ & $0.21$ & $0.20$ & $0.28$ & $0.30$ & $0.37$ & $0.42$ & $0.51$ & $0.60$ & $0.73$ & $0.90$ & $1.10$ & $1.34$ & $1.62$ & $1.91$ & $2.18$ & $2.46$ & $2.67$ & $2.91$\\
		\hline
	\end{tabular}
	\label{tab:parameters}
\end{table*}
\begin{table*}
	\centering
	\caption{Iterative decoding thresholds for the turbo-like code ensemble with $(32,26,4)$ Hamming component codes. All algorithms employ GMI-based post-processing.}
	\begin{tabular}{|c|c|c|c|c|c|c|c|c|c|}
		\hline
		Algorithm & BP & Iterative max-log APP & $p=8$ & $p=7$ & $p=6$ & $p=5$ & $p=4$ & $p=3$ & $p=2$ \\
		\hline
		$\left( E_\text{b}/N_0 \right)^\star$ & $1.47$ & $1.54$ & $1.54$ & $1.55$ & $1.58$ & $1.66$ & $1.72$ & $1.87$ & $2.14$ \\
		\hline
	\end{tabular}
	\label{tab:thresholds}
	\vspace{3mm}
	\hrule
\end{table*}

Recall that \ac{ECPD-GMI} is $n$ times more complex than the original Chase-Pyndiah decoding due to the decoding rule in Step 1 in Section~\ref{subsec:Extrinsic_Chase_Pyndiah_Decoding}. To allow fair comparisons with Chase-Pyndiah decoding, we adapt \ac{ECPD-GMI} as follows. We define \ac{ECPD-GMI} without the decoding rule in Step 1 as \ac{MCPD-GMI}. We will use \ac{MCPD-GMI} to obtain finite-length performance curves in a two-step process. First, for a given \ac{SNR}, we decode a sufficiently large number of product code frames in parallel. In each half-iteration we use \eqref{eq:optimal_parameters} to obtain the optimal parameters $\boldsymbol{\theta}^\star$ for the post-processing function. Second, we estimate the \ac{BER} by applying those optimal parameters. Despite their differences, we hope that iterative decoding thresholds under \ac{ECPD-GMI} accurately predict the finite-length performance under \ac{MCPD-GMI}.

First, we simulate the performance of product codes based on $(64,51,6)$, $(128,113,6)$, and $(256,239,6)$ \ac{eBCH} component codes. The rates of the product codes are $0.635$, $0.779$ and $0.872$, respectively. We choose the number of \ac{LRB} positions for the Chase decoder to be $p=5$ for all codes. Fig.~\ref{fig:num_res_1} shows the \ac{BER} after $20$ half iterations, i.e., $10$ row and column decoding runs. Solid curves correspond to MCPD-GMI, while dashed curves correspond to the original Chase-Pyndiah decoder. For the longest code we observe an increased coding gain of approximately $0.11$ dB at a \ac{BER} of $10^{-4}$.

In Fig.~\ref{fig:num_res_2}, we show BER results a product code with $(256,239,6)$ \ac{eBCH} component codes for different numbers of \ac{LRB} positions $p$. Recall that the complexity of original Chase-Pyndiah decoding as well as \ac{MCPD-GMI} scale exponentially in $p$, since the Chase decoder applies \ac{BDD} to $2^p$ words. Vertical lines show estimated iterative decoding thresholds of corresponding infinite length turbo-like code ensembles under the \ac{ECPD-GMI} obtained via Monte-Carlo \ac{DE}. We use $10^{6}$ samples and $50$ half iterations. The onset of the waterfall region is in good agreement with the results from the Monte-Carlo \ac{DE}. The performance gap between $p=3$ and $p=5$ at a \ac{BER} of $10^{-4}$ is around $0.32$~dB, which is within $0.02$~dB to the $0.34$~dB gap predicted by the corresponding thresholds. Remarkably, the performance of \ac{MCPD-GMI} with $p=4$ approaches original Chase-Pyndiah decoding with $p=5$ for low BERs. Thus, the new GMI-based post-processing can cut complexity of Chase-Pyndiah decoding in half while maintaining performance.

Recall that $\left( \gamma^\star, \delta^\star \right)$ are derived for each \ac{SNR}. Table~\ref{tab:parameters} shows the optimal parameters for the $(256^2,239^2,6^2)$ product code with $p=5$ at an \ac{SNR} of $3.7$~dB. The values of $\gamma^\star$ are between zero and one, which suggests that we overestimate reliability if an alternative codeword exists. This observation is in agreement with the fact that the corresponding formula in \eqref{eq:output} stems from max-log \ac{APP} decoding, a decoder known for overestimating reliability \cite{Land}. Contrary, we have $w_i = d_i$ if there is no alternative codeword for some position $i$. Since $d_i \in \{ -1,+1 \}$, $\delta^\star$ acts itself as reliability value.

Finally, we provide in Table~\ref{tab:thresholds} iterative decoding thresholds for various decoders with GMI-based post-processing. \ac{BP} corresponds to a decoder, which decodes the $(32,26,4)$ Hamming component codes with an optimal \ac{APP} decoder. Recall that in the asymptotic setting \ac{BP} computes optimal marginals. Experimental results show that $\gamma^\star \approx 1$ for GMI-based post-processing of \ac{BP}. When employing a suboptimal max-log \ac{APP} component decoder instead, the threshold degrades by only $0.07$ dB to $1.54$ dB. We remark that the threshold of the same algorithm without GMI-based post-processing increases by approximately $0.5$ dB. The thresholds for \ac{ECPD-GMI} increase as we decrease the parameter $p$. However, $p=8$ and the iterative max-log \ac{APP} algorithm yield identical values for the threshold, up to the second decimal digit. The obtained thresholds are in good agreement with finite-length simulations.

\section{Conclusion} \label{sec:conclusion}

We proposed a framework for optimizing the scaling coefficients for Chase-Pyndiah decoding based on maximizing \ac{GMI}. The parameters found improve upon original Chase-Pyndiah decoding with comparable decoding complexity. Iterative decoding thresholds for an extrinsic version of the algorithm accurately predict the \ac{BER} of the modified decoder.

\section*{Acknowledgment}
The authors thank Gerhard Kramer for helpful comments.

\bibliographystyle{IEEEtran}

\bibliography{IEEEabrv,references}

\enlargethispage{-1.4cm} 

\end{document}